\documentclass[twocolumn,showpacs,preprintnumbers,amsmath,amssymb]{revtex4}

\usepackage{graphicx}
\usepackage{dcolumn}
\usepackage{bm}
\usepackage{ifpdf}
\usepackage{epstopdf}
\usepackage{slashed}
\usepackage{amsfonts}
\usepackage{mathrsfs}
\usepackage{amssymb}
\usepackage{multirow}
\usepackage{CJK}
\usepackage{xcolor}
\usepackage{subfigure,dcolumn}
\usepackage[T2A,T1]{fontenc}
\usepackage[english]{babel}
\usepackage{amsmath}
\usepackage{listings}
\usepackage{booktabs}
\usepackage{youngtab}
\usepackage{float}

\renewcommand{\thetable}{\arabic{table}}

\usepackage{longtable}
\begin{document}

\title{Importance of physical information on the prediction of heavy-ion fusion cross section with machine learning}

\author{Zhilong Li}

\affiliation{College of Physics Science and Technology, Shenyang Normal University, Shenyang 110034, Liaoning, China}
\affiliation{School of Science, Huzhou University, Huzhou 313000, China}

\author{Zepeng Gao}
\affiliation{Sino-French Institute of Nuclear Engineering and Technology, Sun Yat-sen University, Zhuhai 519082, China}
\author{Ling Liu}
\email[Corresponding author, ]{liuling@synu.edu.cn}
\affiliation{College of Physics Science and Technology, Shenyang Normal University, Shenyang 110034, Liaoning, China}
\author{Yongjia Wang}
\email[Corresponding author, ]{wangyongjia@zjhu.edu.cn}
\affiliation{School of Science, Huzhou University, Huzhou 313000, China}

\author{Long Zhu}
\affiliation{Sino-French Institute of Nuclear Engineering and Technology, Sun Yat-sen University, Zhuhai 519082, China}
\author{Qingfeng Li}

\affiliation{School of Science, Huzhou University, Huzhou 313000, China}
\affiliation{Institute of Modern Physics, Chinese Academy of Science, Lanzhou 730000, China}

\date{\today}

\begin{abstract}
In this work, the Light Gradient Boosting Machine (LightGBM), which is a modern decision tree based machine-learning algorithm, is used to study the fusion cross section (CS) of heavy-ion reaction. Several basic quantities (e.g., mass number and proton number of projectile and target) and the CS obtained from phenomenological formula are fed into the LightGBM algorithm to predict the CS. It is found that, on the validation set, the mean absolute error (MAE) which measures the average
magnitude of the absolute difference between $log_{10}$ of the predicted CS and experimental CS is 0.129 by only using the basic quantities as the input, this value is smaller than 0.154 obtained from the empirical coupled channel model. MAE can be further reduced to 0.08 by including an physical-informed input feature. The MAE on the test set (it consists of 280 data points from 18 reaction systems that not included in the training set) is about 0.19 and 0.53 by including and excluding the physical-informed feature, respectively. We further verify the LightGBM predictions by comparing the CS of $^{ 40,48}{\rm Ca }$+$^{78}{\rm Ni}$ obtained from the density-constrained time-dependent Hartree-Fock approach. Our study demonstrates the importance of physical information in predicting fusion cross section of heavy-ion reaction with machine learning.

\end{abstract}
\maketitle

\section{Introduction}

The heavy ion fusion reaction is a process in which two colliding atomic nuclei overcome the fusion barrier and then forms a excited compound nucleus. It has important scientific and applied implications, including deeper understanding of the properties and reactions of atomic nuclei, study of the synthesis and properties of superheavy elements, exploring the origin of elements, reference for the development and application of fusion energy. In addition, it is one of the most important ways for us to explore the boundary of nuclear landscape to get insights into the nuclear interactions. Therefore, it has been a hot topic of research in the field of nuclear physics since 60 years ago \cite{back2014recent,jiang2021heavy,montagnoli2017recent,montagnoli2023recent,sun2022effects}. To perform heavy-ion reactions, a few facilities have been established and some are under construction all over the world, for example, the Cooler-Storage-Ring (CSR) \cite{xia2002heavy} and High Intensity heavy-ion Accelerator Facility (HIAF) in China \cite{yang2013high,zhou2022status}, rare-isotope beam accelerator complex (RAON) in Korea \cite{hong2023status}, the Facility for Rare Isotope Beams (FRIB) in the United States, the Facility for An-tiproton and Ion Research (FAIR) in Gremany, the Système de Production d'Ions Radioactifs (SPIRAL2) in France, the Radioactive Isotope Beam Factory (RIBF) in Japan. By using various facilities, more than 1000 excitation functions for different projectile-target combinations have been measured \cite{karpov2017nrv}, but there are still many that have not been measured, or have large errors.

The heavy-ion fusion reaction is a complex quantum many-body process, and it involves the mutual coupling of nuclear structure and reaction dynamics \cite{frobrich1984fusion,balantekin1998quantum,dasgupta1998measuring,canto2006fusion,diaz2007effects,zhang2011heavy,canto2015recent,cheng2022non}. Thus it is very difficult to study strictly with first principles. Several theoretical models or empirical formulas have been proposed to study the fusion cross section (CS) which is one of the most important observables for studying heavy-ion reactions. Such as the coupled channel calculations \cite{hagino1999program,wen2020near}, the time-dependent Hartree-Fock (TDHF) theory plus solving Schrodinger equation \cite{simenel2018heavy,simenel2012nuclear}, and empirical coupled channel model \cite{wang2017systematics,gautam2022comprehensive,wong1973interaction,lwin2017applicability,PhysRevC.103.024607,hill1953nuclear,zhu2014orientation}. But the cross sections calculated by these methods are not completely compatible with the experimental data.

In recent years, Machine learning (ML) methods have been widely and successfully applied for analyzing data in many branches of science, such as physics (see, e.g., Refs.\;\cite{carleo2019machine,he2023machine,he2023high}). In the field of nuclear physics, ML has shown a strong ability in the study of heavy ion collisions \cite{wang2023machine,li2022determination,pang2018equation,wei2023event,wang2021finding},  properties of
strongly interacting QCD matter \cite{zhou2023exploring,kuttan2021equation,li2023deep,du2022jet}, nuclear spallation and projectile fragmentation reactions \cite{peng2022bayesian,ma2022precise,song2023image}, nuclear fission \cite{PhysRevLett.123.122501,wang2021optimizing,wang2022bayesian,song2023image}, nuclear masses \cite{niu2018nuclear,gao2021machine,ming2022nuclear,wu2021nuclear,zhao2022new,le2023nuclear}, $\beta$-decay half-lives and energy \cite{peng2022beta,gao2023investigation,munoz2023predicting}, $\alpha$-decay\cite{li2022deep}, the charge radius of atomic nuclei \cite{dong2023nuclear,su2023progress,wu2020calculation,dong2022novel}, nuclear density distribution \cite{yang2021taming,shang2022prediction}, and the evaporation residual cross sections for superheavy nuclei \cite{zhao2022improvement}. Recently, a novel artificial intelligence approach has been applied to model cross section data \cite{dell2022modeling}, in which phenomenological formulas for the calculation of CS are derived based on a hybridization of genetic programming and artificial neural networks. The derived phenomenological formulas can qualitatively reproduce the trend but not the absolute value of the CS. ML is a rapidly growing and
flourishing field. Nowadays, a diverse
array of ML algorithm has been developed and continue
to be refined to cover a wide variety of data types and
tasks. It is interesting to study if other ML algorithm can also refine models that used in the calculations of CS, and more importantly, whether physical insights into the heavy-ion fusion reaction can be derived.

The rest of this paper is organized as follows. In Section. II, we introduce the methodology that we use in the present work, including the machine learning algorithm, the dataset and the input features. The CS obtained with machine learning algorithm are discussed in detail in Sect. III. The conclusions are given in Sect. IV.

\section{Methodology}
In the present work, the prediction of heavy-ion fusion cross section is a supervised task and requires a machine learning algorithm, and a set of labelled data with input and output variables. In this section, we introduce briefly these items. 

The machine learning algorithm we used in the present work is the Light Gradient Boosting Machine (LightGBM) which was developed by Microsoft in 2016, it is a gradient boosting framework that uses tree based learning algorithms \cite{NIPS2017_6449f44a}.  LightGBM is becoming
increasingly popular due to its advantages including  (1) faster training speed and higher efficiency, (2) lower memory usage, (3) better accuracy, (4) support of parallel and graphics processing unit learning, and (5) capability of
handling large-scale data. Moreover, LightGBM is a white-box model and has an excellent degree of explainability because of its decision-tree-based nature. This is important for studying a real physical problem, as explainability may improve our knowledge about the
relationship between the input features and the output.  In our previous works, the strong ability of LightGBM to refine nuclear mass models \cite{gao2021machine} and mine physical information has been demonstrated \cite{WANG2022137508,li2021application,li2020application}. Thus it is also
employed in the present work, and parameters in LightGBM are set to their
default values; we have checked that the results are insensitive to
parameters in LightGBM.

\begin{longtable*}{@{\extracolsep{\fill}}lll|lll|lll}
\caption{Training set. }
\label{table1} \\
\hline
System & Data & Energy range & System & Data & Energy range & System & Data & Energy range \\
& points & (MeV) & & pointts & (MeV) & & pointts & (MeV) \\ \hline
\endfirsthead
\multicolumn{9}{c}{{\tablename} \thetable{} -- continued from previous page} \\
\hline
System & Data & Energy range & System & Data & Energy range & System & Data & Energy range \\
& points & (MeV) & & pointts & (MeV) & & pointts & (MeV)  \\ \hline
\endhead
\hline \multicolumn{9}{c}{{Continued on next page}} \\ \hline
\endfoot
\hline \hline
\endlastfoot
$^{ 12}{\rm C }$+$^{ 89}{\rm Y }$    & 9           & 26-41   

&$^{ 19}{\rm F }$+$^{ 93}{\rm Nb}$    & 12           & 43-60          

& $^{ 32}{\rm S }$+$^{184}{\rm W }$    & 7            &  118-144 
\\
$^{ 12}{\rm C }$+$^{ 92}{\rm Zr}$   & 16           & 27-45   

&  $^{ 19}{\rm F }$+$^{139}{\rm La}$    & 13           & 61-115

&$^{ 32}{\rm S }$+$^{208}{\rm Pb}$     &32  &139-184   
\\
$^{ 12}{\rm C }$+$^{144}{\rm Sm}$  & 15           & 41-70  

&$^{ 19}{\rm F }$+$^{208}{\rm Pb}$    & 33           & 73-145

& $^{ 33}{\rm S }$+$^{ 90}{\rm Zr}$    & 13            &  74-97    
\\
$^{ 12}{\rm C }$+$^{152}{\rm Sm}$   & 6          & 42-58

&$^{ 27}{\rm Al}$+$^{ 45}{\rm Sc}$     & 16           & 31-51

&$^{ 33}{\rm S }$+$^{ 91}{\rm Zr}$       & 16       &  72-97    
\\
$^{ 12}{\rm C }$+$^{154}{\rm Sm}$   & 12           & 43-58

& $^{ 19}{\rm F }$+$^{209}{\rm Bi}$   &  10         & 80-116

&  $^{ 33}{\rm S }$+$^{ 92}{\rm Zr}$        & 16    & 72-97     
\\
$^{ 12}{\rm C }$+$^{181}{\rm Ta}$    & 12           & 50-92
& $^{ 23}{\rm Na}$+$^{ 48}{\rm Ti}$    &  11        & 32-46

& $^{ 34}{\rm S }$+$^{ 24}{\rm Mg}$    &   11   &   24-32      
\\
$^{ 12}{\rm C }$+$^{194}{\rm Pt}$    & 10          & 50-69
& $^{ 27}{\rm Al}$+$^{ 70}{\rm Ge}$   &21           &50-60
& $^{ 34}{\rm S }$+$^{ 25}{\rm Mg}$    &   19   &   24-33 
\\
$^{ 12}{\rm C }$+$^{198}{\rm Pt}$    & 10          & 50-69
&$^{ 27}{\rm Al}$+$^{ 72}{\rm Ge}$   & 17          &49-61
& $^{ 34}{\rm S }$+$^{ 26}{\rm Mg}$    &   21   &   24-35 
\\
  $^{ 40}{\rm Ca}$+$^{ 96}{\rm Zr}$   &  56          & 87-113
&$^{ 27}{\rm Al}$+$^{ 73}{\rm Ge}$  &18          &50-62
&$^{ 34}{\rm S }$+$^{ 89}{\rm Y }$  &  32         & 72-92
\\
     $^{ 46}{\rm Ti}$+$^{ 90}{\rm Zr}$  &   10         & 99-120
& $^{ 27}{\rm Al}$+$^{ 74}{\rm Ge}$  & 18           &49-62
&$^{ 34}{\rm S }$+$^{168}{\rm Er}$  &  41          & 111-164
\\
$^{ 12}{\rm C }$+$^{204}{\rm Pb}$    & 12          & 50-85
&$^{ 27}{\rm Al}$+$^{ 76}{\rm Ge}$   & 19          &48-62
& $^{ 35}{\rm Cl}$+$^{ 24}{\rm Mg}$  & 12        & 26-36
\\
$^{ 12}{\rm C }$+$^{206}{\rm Pb}$    & 8          & 54-81
&$^{ 27}{\rm Al}$+$^{197}{\rm Au}$   & 29         &107-151
& $^{ 35}{\rm Cl}$+$^{ 25}{\rm Mg}$  & 12        & 27-38
\\
$^{ 12}{\rm C }$+$^{208}{\rm Pb}$    & 12          & 54-89
&$^{ 28}{\rm Si}$+$^{ 28}{\rm Si}$  & 38          &22-68
& $^{ 35}{\rm Cl}$+$^{ 26}{\rm Mg}$  &  12       & 27-38
\\
$^{ 12}{\rm C }$+$^{237}{\rm Np}$    & 11          & 56-77
& $^{ 28}{\rm Si}$+$^{ 30}{\rm Si}$  &22             &22-49
& $^{ 35}{\rm Cl}$+$^{ 27}{\rm Al}$  & 11   &  30-75
\\
$^{ 12}{\rm C }$+$^{238}{\rm U}$    & 21          & 60-119
&$^{ 28}{\rm Si}$+$^{ 68}{\rm Zn}$   &12 &50-71
& $^{ 35}{\rm Cl}$+$^{ 54}{\rm Cr}$  &  17 &  52-79
\\
$^{ 14}{\rm N }$+$^{ 59}{\rm Co}$     & 8          & 25-45
& $^{ 28}{\rm Si}$+$^{ 90}{\rm Zr}$  & 13      &65-93
&$^{ 35}{\rm Cl}$+$^{ 52}{\rm Cr}$    & 16     & 53-78
\\
$^{ 14}{\rm N }$+$^{232}{\rm Th}$     & 9          & 67-87
& $^{ 28}{\rm Si}$+$^{ 92}{\rm Zr}$  &22         &65-89
& $^{ 35}{\rm Cl}$+$^{ 51}{\rm V }$  & 18        &49-81
\\
$^{ 14}{\rm N }$+$^{238}{\rm U }$   & 17          & 72-138
& $^{ 28}{\rm Si}$+$^{ 94}{\rm Zr}$  &15          & 63-95
& $^{ 35}{\rm Cl}$+$^{ 50}{\rm Ti}$ &22 &47-75
\\
$^{ 15}{\rm N }$+$^{ 56}{\rm Fe}$  & 10          & 25-37
& $^{ 28}{\rm Si}$+$^{ 93}{\rm Nb}$ &13          &68-92
& $^{ 35}{\rm Cl}$+$^{ 54}{\rm Fe}$  &18 &55-82
\\
$^{ 15}{\rm N }$+$^{209}{\rm Bi}$   & 14          & 63-83
& $^{ 28}{\rm Si}$+$^{142}{\rm Ce}$  & 20  &87-121
& $^{ 35}{\rm Cl}$+$^{ 58}{\rm Ni}$ &9 &60-88
\\
$^{ 16}{\rm O }$+$^{ 58}{\rm Ni}$    & 6          & 31-48
&$^{ 28}{\rm Si}$+$^{154}{\rm Sm}$  &21 &90-123
& $^{ 35}{\rm Cl}$+$^{ 60}{\rm Ni}$   & 8 &59-82
\\
$^{ 16}{\rm O }$+$^{ 62}{\rm Ni}$    & 8         & 31-48
&$^{ 28}{\rm Si}$+$^{164}{\rm Er}$   &12 &  108-151
& $^{ 35}{\rm Cl}$+$^{ 62}{\rm Ni}$   &15   & 59-109
\\
$^{ 16}{\rm O }$+$^{ 59}{\rm Co}$    & 4          & 26-40
& $^{ 28}{\rm Si}$+$^{198}{\rm Pt}$ & 11   & 114-167
& $^{ 35}{\rm Cl}$+$^{ 64}{\rm Ni}$  &  9  & 58-84
\\
$^{ 16}{\rm O }$+$^{ 70}{\rm Ge}$    & 26         & 31-49
& $^{ 28}{\rm Si}$+$^{208}{\rm Pb}$ &35 &   120-160 
& $^{ 35}{\rm Cl}$+$^{ 90}{\rm Zr}$ & 7 &  86-119
\\
$^{ 16}{\rm O }$+$^{ 72}{\rm Ge}$    & 29         & 31-52
&$^{ 29}{\rm Al}$+$^{197}{\rm Au}$   & 19         &109-165
& $^{ 35}{\rm Cl}$+$^{ 92}{\rm Zr}$  & 20         &77-98
\\
$^{ 16}{\rm O }$+$^{ 73}{\rm Ge}$   & 27          & 32-53
&$^{ 31}{\rm Al}$+$^{197}{\rm Au}$   & 20         & 106-165
& $^{ 36}{\rm S }$+$^{ 48}{\rm Ca}$ & 25  & 36-61
\\
$^{ 16}{\rm O }$+$^{ 74}{\rm Ge}$    & 29          & 31-56
&$^{ 30}{\rm Si}$+$^{ 30}{\rm Si}$   & 26          &26-50
&$^{ 36}{\rm S }$+$^{ 64}{\rm Ni}$  & 24 & 51-83
\\
$^{ 16}{\rm O }$+$^{ 76}{\rm Ge}$    & 32          & 31-56
&$^{ 30}{\rm Si}$+$^{238}{\rm U }$  &9              &124-170
&$^{ 36}{\rm S }$+$^{ 90}{\rm Zr}$   &28            &72-90
\\
$^{ 16}{\rm O }$+$^{ 92}{\rm Zr}$    & 36          & 37-70
&$^{ 32}{\rm S }$+$^{ 27}{\rm Al}$   &30           &27-61
&$^{ 36}{\rm S }$+$^{ 96}{\rm Zr}$   & 33          &71-88
\\
$^{ 16}{\rm O }$+$^{112}{\rm Cd}$    & 14          & 45-60
&$^{ 32}{\rm S }$+$^{ 24}{\rm Mg}$   & 15          &26-33
&$^{ 36}{\rm S }$+$^{ 92}{\rm Mo}$    & 5          & 76-86
\\
$^{ 16}{\rm O }$+$^{144}{\rm Nd}$     & 13          & 54-81
&$^{ 32}{\rm S }$+$^{ 25}{\rm Mg}$     &19          &24-33
&$^{ 36}{\rm S }$+$^{ 94}{\rm Mo}$   & 8          & 75-87
\\
$^{ 16}{\rm O }$+$^{148}{\rm Nd}$    & 6           & 58-72
&$^{ 32}{\rm S }$+$^{ 26}{\rm Mg}$    & 20     & 24-34
& $^{ 36}{\rm S }$+$^{ 96}{\rm Mo}$   &7        & 71-88
\\
$^{ 16}{\rm O }$+$^{150}{\rm Nd}$    & 7           & 55-73
&$^{ 32}{\rm S }$+$^{ 40}{\rm Ca}$   & 7              & 44-66
&$^{ 36}{\rm S }$+$^{ 98}{\rm Mo}$    & 7     & 75-88
\\
$^{ 16}{\rm O }$+$^{144}{\rm Sm}$    & 28           & 55-91
&$^{ 32}{\rm S }$+$^{ 48}{\rm Ca}$  &21             &35-55
&$^{ 36}{\rm S }$+$^{100}{\rm Mo}$  & 9            & 74-90
\\
$^{ 16}{\rm O }$+$^{147}{\rm Sm}$   & 8           & 55-68
& $^{ 32}{\rm S }$+$^{ 64}{\rm Ni}$ & 13          & 52-76
&$^{ 36}{\rm S }$+$^{100}{\rm Ru}$  & 8          & 77-90
\\
$^{ 16}{\rm O }$+$^{148}{\rm Sm}$    & 25           & 53-75
& $^{ 32}{\rm S }$+$^{ 89}{\rm Y }$ &23             &72-96
& $^{ 36}{\rm S }$+$^{101}{\rm Ru}$ &  8   & 77-90
\\
$^{ 16}{\rm O }$+$^{149}{\rm Sm}$   & 10           & 55-68
&$^{ 32}{\rm S }$+$^{ 90}{\rm Zr}$   & 41         &72-96
& $^{ 36}{\rm S }$+$^{102}{\rm Ru}$  & 7         & 78-89
\\
$^{ 16}{\rm O }$+$^{150}{\rm Sm}$    & 8           & 54-68
&$^{ 32}{\rm S }$+$^{ 94}{\rm Zr}$  & 50     &  70-95
& $^{ 36}{\rm S }$+$^{103}{\rm Rh}$   & 4 & 79-89
\\
$^{ 16}{\rm O }$+$^{152}{\rm Sm}$    & 7           & 54-70
&$^{ 32}{\rm S }$+$^{ 96}{\rm Zr}$  &   50        &69-98
&$^{ 36}{\rm S }$+$^{104}{\rm Pd}$  &  4 &   81-90
\\
$^{ 16}{\rm O }$+$^{154}{\rm Sm}$   & 39           & 52-100
&$^{ 32}{\rm S }$+$^{ 94}{\rm Mo}$   & 9      &76-90
& $^{ 36}{\rm S }$+$^{104}{\rm Ru}$  &  10 & 77-89
\\

$^{ 16}{\rm O }$+$^{166}{\rm Er}$     & 17           & 58-94
&$^{ 32}{\rm S }$+$^{ 96}{\rm Mo}$   &       9 &74-90
& $^{ 36}{\rm S }$+$^{105}{\rm Pd}$  &4  & 81-90
\\
$^{ 16}{\rm O }$+$^{176}{\rm Yb}$   & 15           & 59-89
& $^{ 32}{\rm S }$+$^{ 98}{\rm Mo}$  &10           &73-91
& $^{ 36}{\rm S }$+$^{106}{\rm Pd}$  &5 &  80-90
\\
$^{ 16}{\rm O }$+$^{186}{\rm W }$    & 35            & 62-83
& $^{ 32}{\rm S }$+$^{ 100}{\rm Mo}$  &11             & 72-91
&$^{ 36}{\rm S }$+$^{108}{\rm Pd}$   &6               &79-95
\\
$^{ 16}{\rm O }$+$^{204}{\rm Pb}$    & 12           & 66-98
& $^{ 32}{\rm S }$+$^{100}{\rm Ru}$  &8     & 77-91
& $^{ 36}{\rm S }$+$^{110}{\rm Pd}$ & 6 &79-95
\\
$^{ 16}{\rm O }$+$^{208}{\rm Pb}$    & 38           & 65-110
&$^{ 32}{\rm S }$+$^{101}{\rm Ru}$   &6     &  80-91
& $^{ 36}{\rm S }$+$^{204}{\rm Pb}$ & 26    &138-179
\\
$^{ 16}{\rm O }$+$^{209}{\rm Bi}$    & 14           & 71-95
&$^{ 32}{\rm S }$+$^{102}{\rm Ru}$   & 9            &76-92
&$^{ 36}{\rm S }$+$^{238}{\rm U }$  & 5  &147-173
\\
$^{ 16}{\rm O }$+$^{232}{\rm Th}$    & 12           & 70-150
&$^{ 32}{\rm S }$+$^{103}{\rm Rh}$   & 8            &78-92
& $^{37}{\rm Cl}$+$^{ 24}{\rm Mg}$ &12             & 25-35
\\
$^{ 16}{\rm O }$+$^{238}{\rm U }$   & 14           & 80-157
& $^{ 32}{\rm S }$+$^{104}{\rm Pd}$ & 6            & 81-92
& $^{ 37}{\rm Cl}$+$^{ 25}{\rm Mg}$ & 12           & 26-36
\\
$^{ 17}{\rm O }$+$^{144}{\rm Sm}$   & 25           & 54-90
& $^{ 32}{\rm S }$+$^{104}{\rm Ru}$ & 9       & 77-92
& $^{ 37}{\rm Cl}$+$^{ 26}{\rm Mg}$  &  12    & 26-38
\\
$^{ 18}{\rm O }$+$^{ 44}{\rm Ca}$   & 17           & 19-43
& $^{ 32}{\rm S }$+$^{105}{\rm Pd}$ &5           &81-92
&$^{ 37}{\rm Cl}$+$^{ 59}{\rm Co}$   &13         & 55-71
\\
$^{ 18}{\rm O }$+$^{ 74}{\rm Ge}$    & 26           & 30-50
& $^{ 32}{\rm S }$+$^{106}{\rm Pd}$  & 7           & 80-92
&$^{ 37}{\rm Cl}$+$^{ 70}{\rm Ge}$  &      12 &  62-76

\\
$^{ 18}{\rm O }$+$^{112}{\rm Sn}$    & 17           & 42-78
& $^{ 32}{\rm S }$+$^{108}{\rm Pd}$   & 8            &79-92
&$^{ 37}{\rm Cl}$+$^{ 72}{\rm Ge}$   &  14  & 63-77
\\
$^{ 18}{\rm O }$+$^{118}{\rm Sn}$     & 17           & 44-79
& $^{ 32}{\rm S }$+$^{110}{\rm Pd}$   & 9            &78-93
& $^{ 37}{\rm Cl}$+$^{ 73}{\rm Ge}$  & 12    &63-77
\\
$^{ 18}{\rm O }$+$^{124}{\rm Sn}$    & 17           & 44-79
&$^{ 32}{\rm S }$+$^{138}{\rm Ba}$   &  13          & 99-135
& $^{ 37}{\rm Cl}$+$^{ 74}{\rm Ge}$  & 9            & 63-75
\\
$^{ 18}{\rm O }$+$^{192}{\rm Os}$    & 7           & 73-113
&$^{ 32}{\rm S }$+$^{154}{\rm Sm}$   & 9           &101-128
&  $^{ 37}{\rm Cl}$+$^{ 76}{\rm Ge}$ &  14         & 62-76
\\
$^{ 18}{\rm O }$+$^{208}{\rm Pb}$   & 18           & 68-95
& $^{ 32}{\rm S }$+$^{182}{\rm W }$ &  15          & 122-170
& $^{ 37}{\rm Cl}$+$^{ 93}{\rm Nb}$  & 7           & 81-93
\\
 $^{ 37}{\rm Cl}$+$^{ 98}{\rm Mo}$   & 7           & 82-95
& $^{ 40}{\rm Ca}$+$^{194}{\rm Pt}$  &  31          & 159-200
& $^{ 58}{\rm Ni}$+$^{ 54}{\rm Fe}$  & 25           & 85-110
\\
  $^{ 37}{\rm Cl}$+$^{ 100}{\rm Mo}$  & 8            & 83-95
&  $^{ 40}{\rm Ca}$+$^{197}{\rm Au}$ &  10          & 162-223
&  $^{ 58}{\rm Ni}$+$^{ 58}{\rm Ni}$ &   17        & 95-109
\\
 $^{ 40}{\rm Ar}$+$^{112}{\rm Sn}$  & 15           & 96-126
&$^{ 40}{\rm Ca}$+$^{208}{\rm Pb}$  &  5           & 169-208
& $^{ 58}{\rm Ni}$+$^{ 64}{\rm Ni}$   &  27          & 89-110
\\
 $^{ 40}{\rm Ar}$+$^{116}{\rm Sn}$   & 14           & 95-128
& $^{ 48}{\rm Ca}$+$^{ 48}{\rm Ca}$ &  29          & 46-63
&  $^{ 58}{\rm Ni}$+$^{124}{\rm Sn}$  &  6          & 143-162
\\
 $^{ 40}{\rm Ar}$+$^{122}{\rm Sn}$   &  17          & 94-130
& $^{ 48}{\rm Ca}$+$^{ 90}{\rm Zr}$  &  33          & 90-115
& $^{ 64}{\rm Ni}$+$^{ 64}{\rm Ni}$  &  13          & 89-107
\\
  $^{ 40}{\rm Ar}$+$^{144}{\rm Sm}$  &  11          & 116-140
& $^{ 48}{\rm Ca}$+$^{ 96}{\rm Zr}$  &    38        & 88-113
& $^{ 64}{\rm Ni}$+$^{124}{\rm Sn}$  &  10          & 144-170
\\
  $^{ 40}{\rm Ar}$+$^{148}{\rm Sm}$  &  12          & 112-150
&$^{ 48}{\rm Ca}$+$^{154}{\rm Sm}$   &   51         & 125-192
&  $^{124}{\rm Sn}$+$^{ 40}{\rm Ca}$ &  22          & 107-150
\\
 $^{ 40}{\rm Ar}$+$^{154}{\rm Sm}$   &  12          & 108-145
& $^{ 48}{\rm Ca}$+$^{197}{\rm Au}$  &   8         & 164-245
& $^{132}{\rm Sn}$+$^{ 40}{\rm Ca}$  &  15          & 109-136
\\
  $^{ 40}{\rm Ca}$+$^{ 40}{\rm Ca}$ &    15         & 49-70
& $^{ 48}{\rm Ca}$+$^{208}{\rm Pb}$ &   5         & 171-206
&  $^{134}{\rm Te}$+$^{ 40}{\rm Ca}$  &    10        & 115-140
\\
 $^{ 40}{\rm Ca}$+$^{ 44}{\rm Ca}$  &     15        & 47-75
& $^{ 48}{\rm Ti}$+$^{ 58}{\rm Fe}$  &   27         & 64-85
&  $^{124}{\rm Sn}$+$^{ 48}{\rm Ca}$ &  13          & 109-170
\\
 $^{ 40}{\rm Ca}$+$^{ 46}{\rm Ti}$  &  25          & 54-80
& $^{ 48}{\rm Ti}$+$^{ 58}{\rm Ni}$  &    14        & 72-91
&  $^{132}{\rm Sn}$+$^{ 48}{\rm Ca}$  & 12           & 109-170
\\
 $^{ 40}{\rm Ca}$+$^{ 48}{\rm Ca}$  & 14           & 48-67
&$^{ 48}{\rm Ti}$+$^{ 60}{\rm Ni}$   &   15         & 72-93
& $^{132}{\rm Sn}$+$^{ 58}{\rm Ni}$   &  10          & 152-200
\\
 $^{ 40}{\rm Ca}$+$^{ 48}{\rm Ti}$  &  28          & 53-82
& $^{ 46}{\rm Ti}$+$^{ 64}{\rm Ni}$  &  15          & 71-96
& $^{132}{\rm Sn}$+$^{ 64}{\rm Ni}$   &  9          & 147-200
\\
 $^{ 40}{\rm Ca}$+$^{ 50}{\rm Ti}$  &   30         & 53-83
& $^{ 48}{\rm Ti}$+$^{ 64}{\rm Ni}$  &    18        & 71-96
& $^{208}{\rm Pb}$+$^{ 26}{\rm Mg}$  &     7       & 109-190
\\
 $^{ 40}{\rm Ca}$+$^{ 58}{\rm Ni}$  &   21        & 67-95
&$^{ 48}{\rm Ti}$+$^{122}{\rm Sn}$   &   19         & 121-160
& $^{ 50}{\rm Ti}$+$^{ 90}{\rm Zr}$  &   8         & 100-118
\\
$^{ 40}{\rm Ca}$+$^{ 64}{\rm Ni}$   &   26         & 63-94
& $^{ 50}{\rm Ti}$+$^{ 60}{\rm Ni}$  &  14          & 72-91
&  $^{ 40}{\rm Ca}$+$^{192}{\rm Os}$    &  35          & 151-200
\\
 $^{ 40}{\rm Ca}$+$^{ 90}{\rm Zr}$   &   40         & 90-120
& $^{ 46}{\rm Ti}$+$^{ 93}{\rm Nb}$  &   10         & 100-120

\\
\cline{7-9}
 $^{ 40}{\rm Ca}$+$^{ 94}{\rm Zr}$  &   59         & 88-115
& $^{ 50}{\rm Ti}$+$^{ 93}{\rm Nb}$  &   9         & 100-120
&Total systems & 220\\
& & & & & &Total points & 3635\\
\end{longtable*}

\begin{table}[!htb]
\caption{Test set.}
\label{table2}
\begin{tabular*}{8.5cm}{@{\extracolsep{\fill}}ccccccccc}
\toprule
System & Data & energy range    & Ref.  \\
        &points          &(MeV)   &           &     \\
\midrule

$^{35}$Cl + $^{130}$Te 
& 16              & 90-125     &\cite{sahoo2020role} 
\\
$^{37}$Cl + $^{130}$Te 
& 17              & 90-125 & \cite{sahoo2019sub}          \\
$^{37}$Cl + $^{68}$Zn  
& 11              & 60-90   &\cite{chauhan2020evaporation}        \\
$^{16}$O + $^{61}$Ni  
& 8              & 25-41   &\cite{deb2022investigation}          \\
$^{18}$O + $^{61}$Ni  
& 16              & 25-41   &\cite{deb2022investigation}         \\
$^{18}$O + $^{62}$Ni  
& 15              & 25-42 &\cite{deb2022investigation}              \\
$^{18}$O + $^{116}$Sn  
& 21              & 40-75 &\cite{kalita2021role} 
\\
$^{30}$Si + $^{156}$Gd  
& 14              & 90-116    &\cite{prajapat2022fusion}         \\

$^{28}$Si + $^{100}$Mo  
& 13              & 65-98    &\cite{stefanini2021new}   \\
$^{36}$S + $^{50}$Ti  
& 23              & 40-60    &\cite{colucci2019sub}   \\
$^{36}$S + $^{51}$V  
& 22              & 40-60    &\cite{colucci2019sub}   \\

$^{12}$C + $^{182}$W  
& 14              & 40-80    &\cite{sanila2022fusion}   \\
$^{12}$C + $^{184}$W 
& 14              & 40-80    &\cite{sanila2022fusion}   \\
$^{12}$C + $^{186}$W 
& 14              & 40-80    &\cite{sanila2022fusion}   \\
$^{40}$Ca + $^{92}$Zr  
& 16              & 89-108    &\cite{stefanini2017new}   \\

$^{48}$Ca + $^{116}$Cd  
& 20              & 104-130    &\cite{del2023influence}   \\
$^{48}$Ca + $^{118}$Sn  
& 16              & 104-130    &\cite{del2023influence}   \\
$^{48}$Ca + $^{120}$Te  
& 10              & 104-130    &\cite{del2023influence}   \\

\hline
Total systems &18\\
Total points & 280
\\
\toprule
\end{tabular*}
\end{table}
Data used in this work consists of three parts, the training set, the validation set, and test set. Training set is used to adjust the parameters in LightGBM, usually validation set is a part of training set which is used to monitor and avoid overfitting. Test set is used
to evaluate the actual predictive power of LightGBM on unseen data. The training set is made of 3635 experimental data points from 220 reaction systems collected in Ref.\cite{wang2017systematics}, it will be randomly split into the training set and validation set with a certain ratio. These experimental data were measured before 2016. We note that the training set is built considering systems with $12 \leq Z_1 \leq 48$ and $24 \leq Z_2 \leq 208$. In this way, we can neglect too heavy systems, for which fusion-fission and quasi-fission can be the dominant reaction modes, and also too light systems, where the presence of break-up and transfer reactions complicate the analysis.  The test set consists of 280 data points from 18 reaction systems measured in recent experiments \cite{sahoo2020role,sahoo2019sub,chauhan2020evaporation,deb2022investigation,kalita2021role,prajapat2022fusion,stefanini2021new,colucci2019sub,sanila2022fusion,stefanini2017new,del2023influence}. We note here that these 18 reaction systems in the test set have not appeared in the training set. The reaction system, number of data points for each system, and the energy range of the training and test sets are listed in Tab.\ref{table1} and \ref{table2}, respectively.

Usually, fusion cross sections at different energies varying over several orders of magnitude, thus the logarithm of fusion cross section is set as the output of LightGBM. The input quantities, including center-of-mass energy, the charge, neutron, mass numbers, and binding energy of projectile, target, and the compound nuclei, the fusion Q-value, as well as the one (two)-proton (neutron) separation energies of the compound nuclei, are listed in Tab.\ref{table3}. These quantities are chosen because they are basic features (BF) of a nucleus and expected to relate to the fusion process. Usually, adding more related features may benefit the performance of the trained ML model. We have tried to add more features and different combinations of these BF quantities, but the performance of the trained ML model is only slightly improved. To introduce physical information related to heavy-ion fusion process, the fusion cross sections calculated with the empirical coupled channel (ECC) model and Wong formula are also used. Details of ECC model and the Wong formula can be found in Ref. \cite{wang2017systematics}. In addition, we introduce a simplified empirical quantity Z$_1$Z$_2$/E$_{c.m}$. This quantity includes some important factors of heavy-ion fusion reaction, e.g., Z$_1$Z$_2$ relates to the Coulomb barrier. These quantities are physical-informed or physical-guided features, which are listed in Tab.\ref{table4}.

\begin{table}[!htb]
\begingroup
\setlength{\tabcolsep}{16pt} 
\renewcommand{\arraystretch}{1} 
\caption{Selection of basic features (BF).}
\label{table3}
\begin{tabular*}{8.7cm} {lllllllll}
\toprule
 \hline
 features & Description \\
\hline
E$\rm _{c.m}$  &  collision center-of-mass energy (MeV) \\
Z$_1$   &  charge of the first reaction partner      \\
         
N$_1$   &  number of neutrons of the first reaction    \\
          &  partner\\
A$_1$  & mass number of the first reaction partner \\
      
Z$_2$   &  charge of the second reaction partner      \\
         
N$_2$   &  number of neutrons of the second reaction    \\
          &  partner\\
A$_2$  & mass number of the second reaction partner \\
Z$_3$   & charge of the compound nucleus      \\
         
N$_3$   &  number of neutrons of the compound     \\
          &  nucleus \\
A$_3$  & mass number of the compound nucleus  \\
B$_1$   & binding energy of the first reaction partner      \\
         
B$_2$   & binding energy of the second reaction  \\
          &   partner\\
B$_3$  & binding energy of the compound nucleus  \\
Q  & fusion Q-value (MeV)     \\
         
S$\rm _p$   & one-proton separation energy of the   \\
          &   compound nucleus (MeV)\\
S$\rm _{2p}$  & two-proton separation energy of the  \\
       &compound nucleus (MeV)\\
S$\rm _n$   & one-neutron separation energy of the  \\
          &   compound nucleus (MeV)\\
S$_{2n}$  & two-neutron separation energy of the  \\
       &compound nucleus (MeV)\\

\bottomrule
\end{tabular*}
\endgroup
\end{table}
\begin{table}[!htb]
\begingroup
\setlength{\tabcolsep}{16pt} 
\renewcommand{\arraystretch}{1} 
\caption{ Physical-informed quantities}
\label{table4}
\begin{tabular*}{8.5cm} {ll}
\toprule
 \hline
features & Description \\
 \hline
sig\_ECC & $\log_{10}$ of CS obtained by ECC model \\
sig\_W &$\log_{10}$ of CS obtained by
Wong formula \\
sig\_E  & Z$_1$Z$_2$/E$_{c.m}$\\
\bottomrule
\end{tabular*}
\endgroup
\end{table}

\begin{table}[!htb]
\begingroup
\setlength{\tabcolsep}{16pt} 
\renewcommand{\arraystretch}{1} 
\caption{ Different modes with different input features.}
\label{table5}
\begin{tabular*}{8.4cm} {ll}
\toprule
 \hline
Mode name & Input features\\
 \hline
Mode\_BF &BF\\
Mode\_ECC &BF+sig\_ECC  \\
Mode\_W &BF+sig\_W\\
Mode\_E  &BF+sig\_E  \\
\bottomrule
\end{tabular*}
\endgroup
\end{table}

The main aim of this work is to establish the relationship between the characteristic quantities of fusion reaction and the fusion cross section by learning the training set with LightGBM. In the process of training, four different input feature combinations are used and given in Tab.\ref{table5}. Mode\_BF represents the input feature is comprised of the 18 basic quantities as listed in Tab.\ref{table3}. Mode\_ECC represents the input feature is comprised of the 18 basic quantities and sig\_ECC. Mode\_W and Mode\_E represent the input
features are comprised of the 18 basic quantities as well as, sig\_W and sig\_E, respectively. By comparing the performances of these different modes, one can infer the importance of physical-informed features. The performance of ML algorithm can be quantitatively evaluated via the mean absolute error (MAE),
\begin{equation}\label{eq2}
\text { MAE }=\frac{1}{N} \sum_{i=1}^{N} \mid \log_{10}\left(\sigma_ {pred }\right)-\log_{10}\left(\sigma_{exp }\right) \mid
\end{equation}
Here N is the number of tested data points, $\sigma_ {pred}$ and $\sigma_ {exp}$ are the predicted and experimental cross section, respectively. 

\section{Results}

\subsection{Performance on the training set}

\subsubsection{Effect of the size of training set}

\begin{figure}[hbt!]
    \centering
    \includegraphics[width=\linewidth]{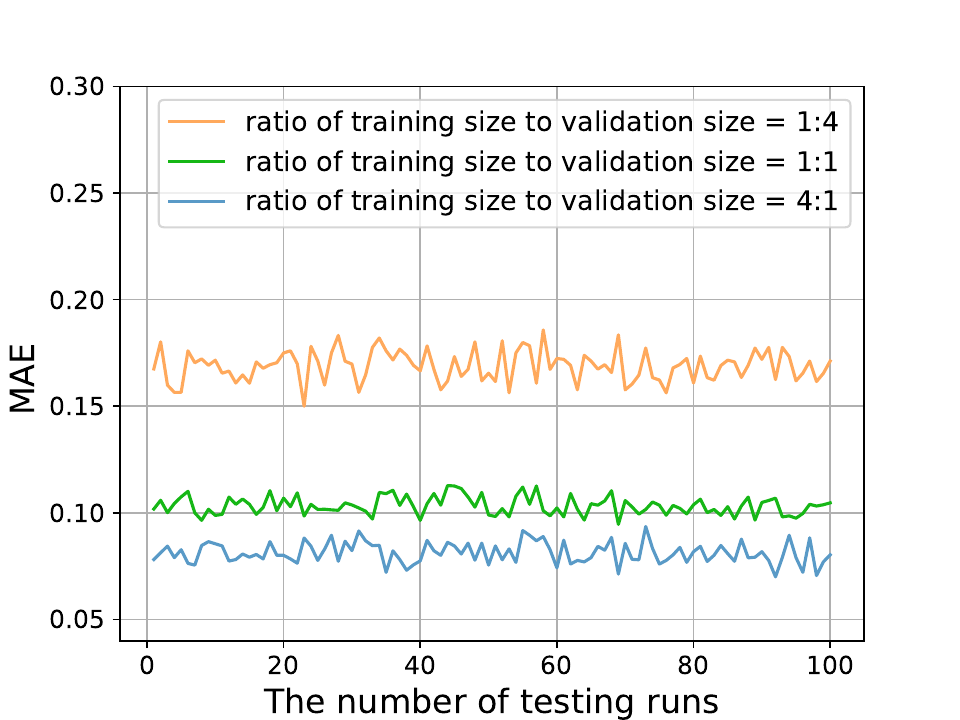}
    \caption{ (Color online) MAE on the validation data obtained with Mode\_E from 100 runs. In each run, the 3635 data points were randomly split into training and validation data sets at a ratio of 4:1 (blue), 1:1 (green), and 1:4 (orange)}
    \label{fig:1}
\end{figure}

\begin{figure}[hbt!]
    \centering
    \includegraphics[width=\linewidth]{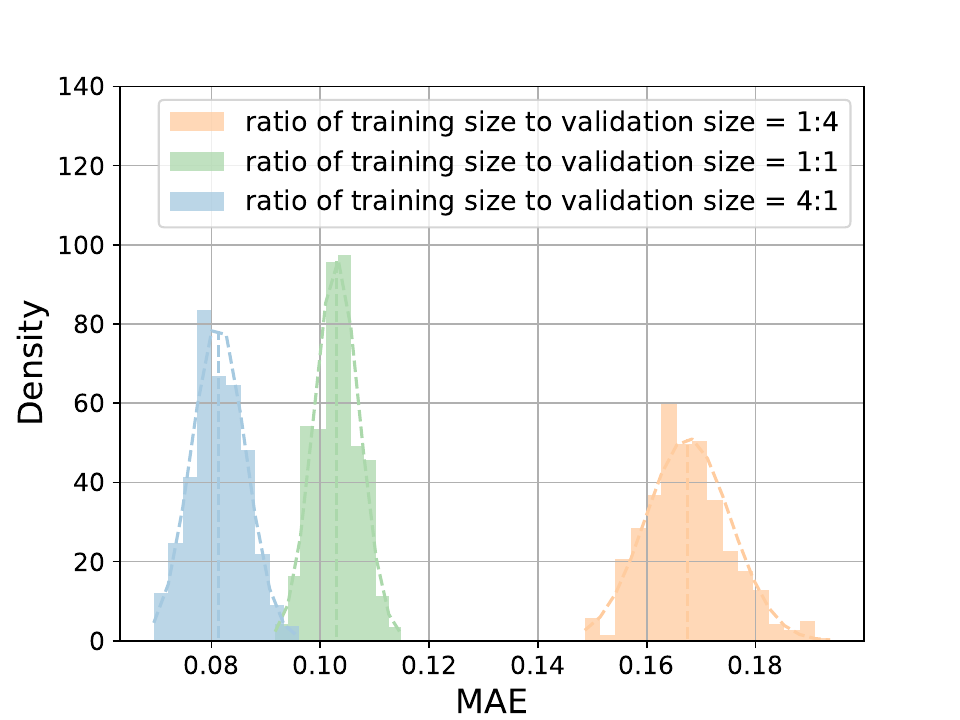}
    \caption{ (Color online) Density distribution of MAE on the validation set. The results from 500 runs for each set are displayed. Dashed lines denote a Gaussian fit to the
distribution. The mean values and the standard deviation of MAE values are 0.167, 0.103 and 0.081, and 0.008, 0.004 and 0.005
for the three sets with different ratios of training to test, respectively.}
    \label{fig:2}
\end{figure}

\begin{figure}[hbt!]
    \centering
    \includegraphics[width=\linewidth]{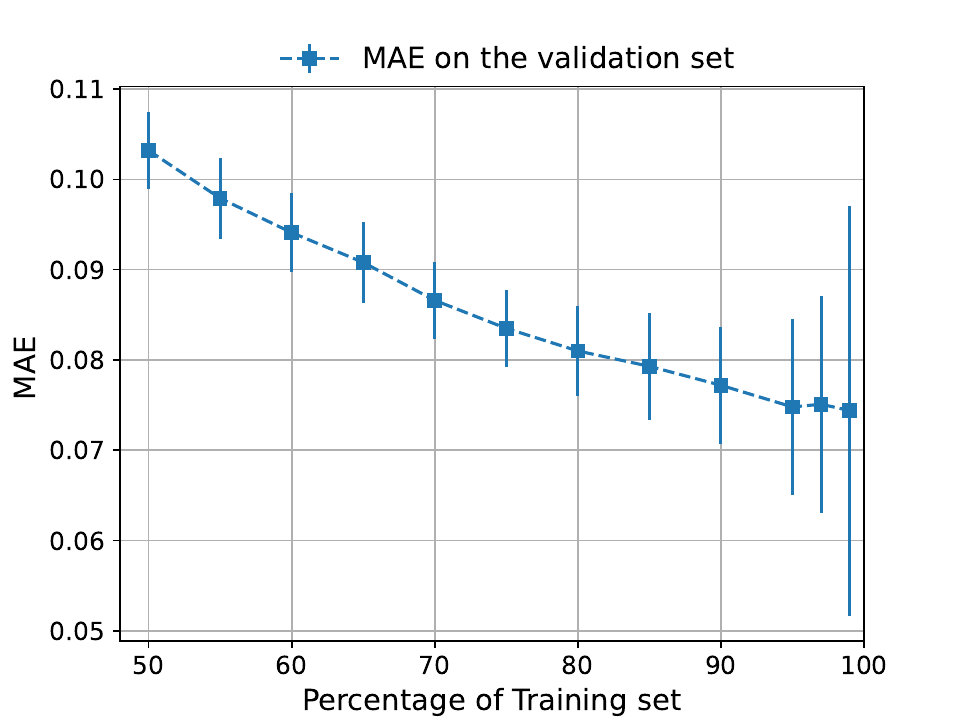}
     \caption{ (Color online) MAE on the validation data obtained with Mode\_E is plotted
as a function of the percentage of the training set.}
    \label{fig:3}
\end{figure}
In this section, the effect of the ratio between the training set and the validation set on the performance of LightGMB in the Mode\_E is studied. For this purpose, the 3635 data points listed in Table \ref{table1} are randomly split into training and validation sets 500 times with each ratio (i.e., 1:4, 1:1, and 4:1). The MAE and
its density distribution are plotted in Figs. \ref{fig:1} and \ref{fig:2}. As observed in Fig. \ref{fig:1}, the MAE is the largest of all when the ratio of training size to validation size is 1:4. The
MAE for 2908 (the validation set, i.e., 80\% of the whole training set) data points predicted by LightGBM with 727 data points was 0.167$\pm$0.041, which means LightGBM can reproduce experimental data within a factor of 10$^{0.167}$=1.47. This is comparable to some physical models. With the training data set built from 2908 data points
and the remaining 727 data points constitute the validation data set, the MAE reduces to 0.081$\pm$0.005, which is better than that of many physical models. The value of MAE fluctuates in different runs, this is because when the training and validation data sets
are randomly selected, it may have a probability that all data points from one reaction system are not included in the training set, the prediction of CS for this system is a challenging task, and resulting in a larger MAE. If a few data points of a reaction system are selected in the training set, the prediction of remaining data on the excitation function of this reaction system is much easier than that on a excitation function without any data point.

By increasing the capacity of the training set, the model can learn more information and reduce the MAE. However, the uncertainty of MAE on the validation set increases with an increase in the percentage of the training set, because the less tested data points, the larger the fluctuations, as shown in Fig. \ref{fig:3}.  In the present work, to avoid either a large MAE value or large uncertainty of MAE, the ratio between the training size and validation size is chosen as 4:1 in the following discussion.

\subsubsection{Comparison of different modes}

\begin{figure}[hbt!]
    \centering
    \includegraphics[width=0.9\linewidth]{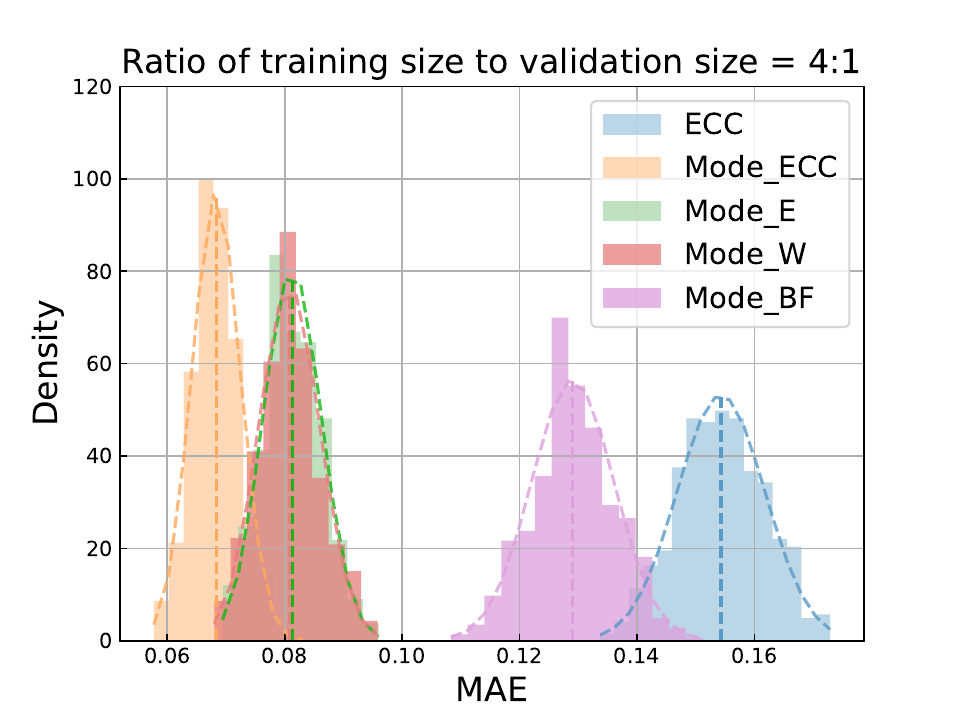}
    \caption{(Color online) Density distribution of MAE for different modes. Results from 500 runs for each mode (Mode\_ECC, Mode\_E, and Mode\_W) and from ECC model are displayed. Dashed lines denote a Gaussian fit to the distribution. In each run, the 3635 fusion reactions were randomly split into training and test sets at a ratio of 4:1  }
    \label{fig:4}
\end{figure}
\begin{table}[!htb]
\begingroup
\setlength{\tabcolsep}{16pt} 
\renewcommand{\arraystretch}{1} 
\caption{The average MAE on the validation set obtained from different modes and from Wong formula and ECC model. The ratio 
of training set to test set is 4:1.}
\label{table6}
\begin{tabular*}{8.7cm} {ll}
\toprule
Model & MAE \\
 \hline
Wong Formula         &      2.38 $\pm$ 0.09\\
ECC model          &     0.154 $\pm$ 0.008\\
Mode\_BF     &      0.129 $\pm$ 0.007\\
Mode\_ECC    &      0.068 $\pm$ 0.004  \\
Mode\_W      &      0.081 $\pm$ 0.005\\
Mode\_E      &      0.081 $\pm$ 0.005\\
\bottomrule
\end{tabular*}
\endgroup
\end{table}

This section aims at demonstrating the impact of input features on the performance of LightGBM. To do so, the density distribution of MAE for Mode\_E, Mode\_BF, Mode\_ECC, and Mode\_W, together with the MAE from ECC model are displayed Fig.\ref{fig:4}. The corresponding mean values and standard deviations are listed in Table \ref{table6}.

First, the MAE for Wong formula is about 2.381, which means the average difference between the predicted CS and experimental CS are as large as two orders of magnitude. This is understandable, as the training set consists of many data points at deep sub-barrier energies, where Wong formula is unreliable.

Second, MAE for Mode\_BF is 0.129 which is smaller than that obtained with ECC model. ECC model includes the effects of neutron transfer channels, the couplings between the relative motion and intrinsic degrees of freedom, as well as the nuclear deformation, it can give a reasonable fit to the fusion excitation function in
the vicinity of the Coulomb barrier. By feeding the basic features of a reaction, LightGBM is able to achieve a better performance on the prediction of CS than ECC model.

Third, the value of MAE can be significantly reduced by including the physical-informed features in the input, this manifests the importance of physical information on the prediction of CS with ML algorithm. MAE for Mode\_ECC is 0.068$\pm$0.004, which is the smallest of all. The values of MAE for Mode\_W and Mode\_E are slightly larger than that of Mode\_ECC. Considering the fact that the calculations of CS with Wong formula or ECC model are much complicated than Z$_1$Z$_2$/E$_{c.m}$, Mode\_E is much favored over other modes.

\subsection{Performance on the test set}

\begin{figure*}[hbt!]
    \centering
    \includegraphics[width=0.9\linewidth]{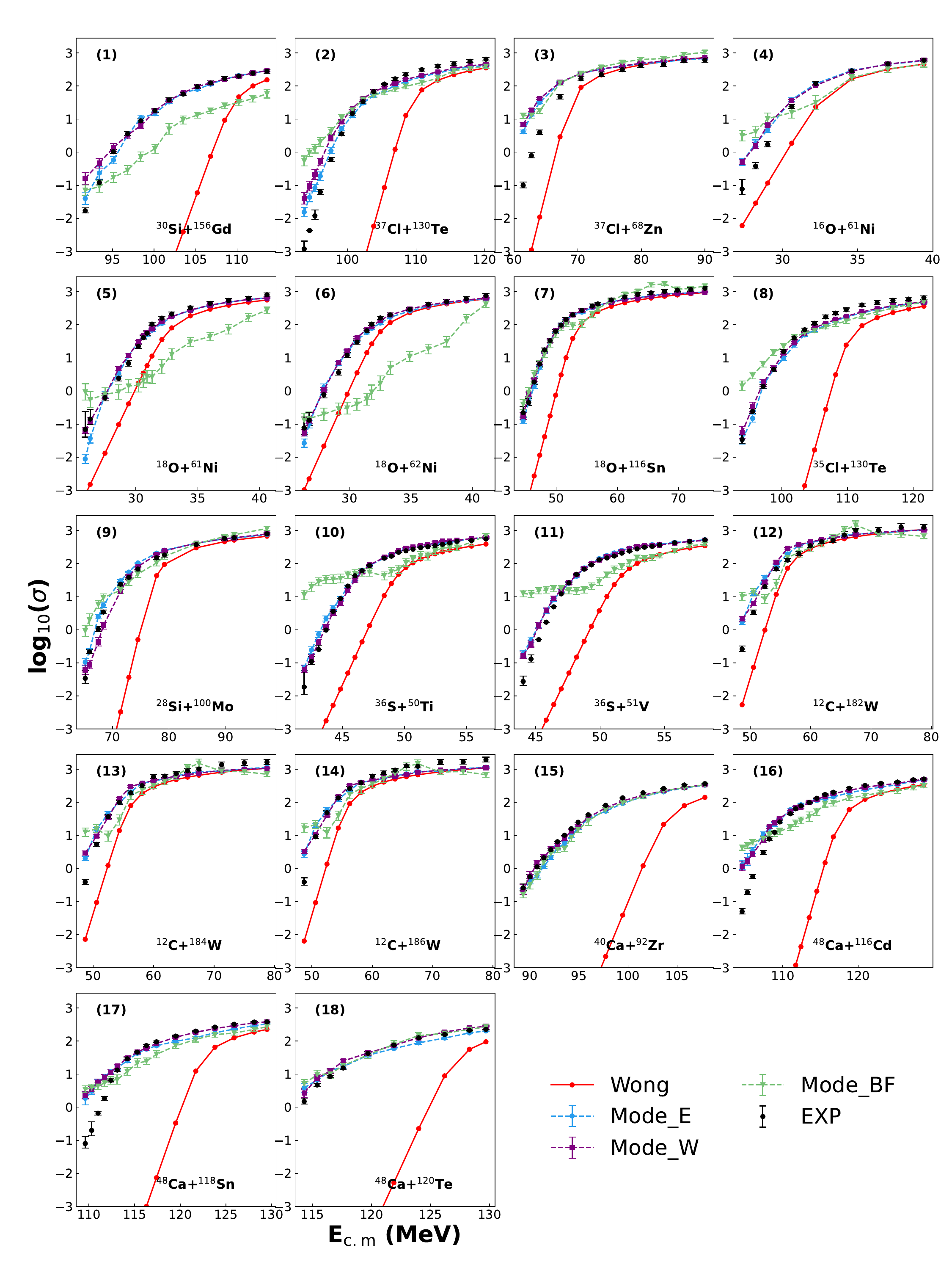}
    \caption{(Color online) The nuclear fusion cross sections predicted by different modes and by Wong formula. Black dots represent experimental data taken from Refs. \cite{sahoo2020role,sahoo2019sub,chauhan2020evaporation,deb2022investigation,kalita2021role,prajapat2022fusion,stefanini2021new,colucci2019sub,sanila2022fusion,stefanini2017new,del2023influence}.}
    \label{fig:5}
\end{figure*}

The performance of LightGBM is further validated using the test set which consists of 280 data points from 18 reaction systems. Fig.\ref{fig:5} displays the comparison of the CS predicted by Mode\_BF, Mode\_E, and Mode\_W with the recent experimental data. It can be seen that the CS obtained with Wong formula is close to experimental data only at high energies, and is too small at low energies. This phenomenon is caused by the invalidation of the parabolic approximation at energies well below the Coulomb barrier, has been widely found and discussed in literature, see e.g., Refs. \cite{hagino2012subbarrier,jiang2021heavy}. Both the CS and its energy-dependent behaviour predicted by Mode\_E and Mode\_W are close to the experimental data, while in some reaction systems, the energy-dependent behaviour obtained by Mode\_BF is very different compared to the  experimental data. The values of MAE on the test set obtained with Mode\_E, Mode\_W, and Mode\_BF are 0.197$\pm$0.006, 0.187$\pm$0.005, and 0.526$\pm$0.013, respectively. This indicates that the physical-informed features can guide machine learning algorithms for successfully capturing the energy-dependent behaviour, then improve the performance. The MAE values on the test set are larger than that on the validation set, this is understandable because reaction systems in the test set are not included in the training set.

\subsection{Comparison with the DC-TDHF approach }

\begin{figure}[hbt!]
    \centering
    \includegraphics[width=0.9\linewidth]{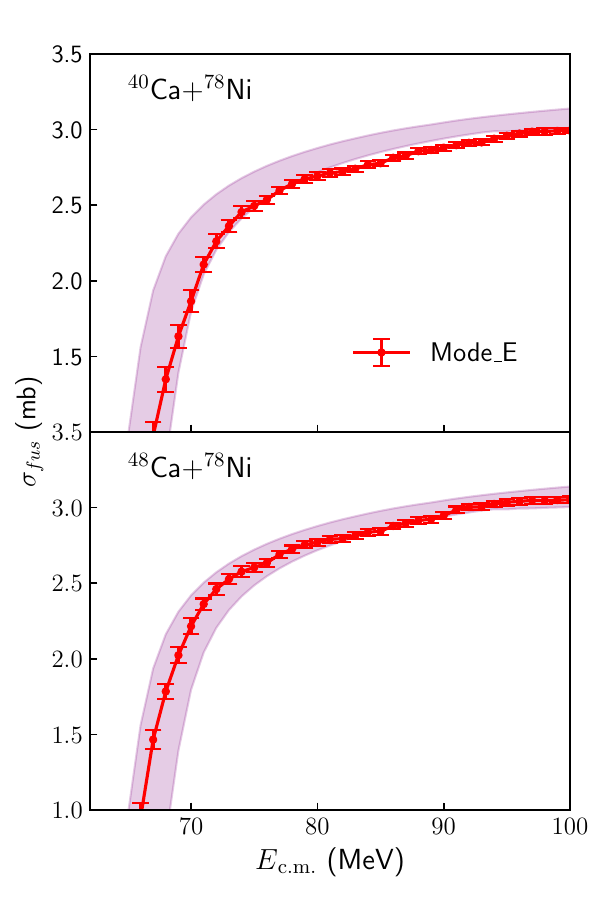}
    \caption{(Color online) The nuclear fusion cross sections for $^{40}$Ca + $^{78}$Ni (upper panel) and $^{48}$Ca + $^{78}$Ni (lower panel). Red points denote the predictions with Mode\_E. The shaded bands denote the calculated results from the density-constrained time-dependent Hartree-Fock (TDHF) approach, taken from Ref. \cite{PhysRevC.105.034601}.}
    \label{fig:6}
\end{figure}

The density-constrained (DC) time-dependent Hartree-Fock (TDHF) is a fully microscopic approach, it provides a good description of the fusion excitation function for many reaction systems \cite{umar2009density,simenel2018heavy}. To further verify the performance of LightGBM, the nuclear fusion cross sections for $^{40,48}$Ca + $^{78}$Ni obtained from DC-TDHF approach are compared with the prediction of Mode\_E, as shown in Fig.\ref{fig:6}. The uncertainties of DC-TDHF result from different potentials \cite{PhysRevC.105.034601}. It can be seen that the results predicted with Mode\_E are in line with the DC-TDHF calculations. However, in contrary to the observed enhancement of
fusion cross sections of $^{40}$Ca + $^{78}$Ni at subbarrier energies in the DC-TDHF calculations, the CS of $^{40}$Ca + $^{78}$Ni predicted with Mode\_E is smaller than that of $^{48}$Ca + $^{78}$Ni. As discussed in Ref.\cite{PhysRevC.105.034601}, this enhancement for $^{40}$Ca + $^{78}$Ni is due to its narrower width of the
ion-ion potential. We note in Refs.\cite{bourgin2016microscopic,bourgin2014barrier} that Bourgin et al. reported the fusion cross section of $^{40}$Ca + $^{64}$Ni system is higher than that of $^{40}$Ca + $^{58}$Ni, because the large neutron transfer probabilities in $^{40}$Ca + $^{64}$Ni result in a lowering of the fusion threshold. The experimental data of $^{40}$Ca + $^{58,64}$Ni are contained in the training set, thus LightGBM should learned from these data and predicted a higher CS for $^{48}$Ca + $^{78}$Ni. Further studies regarding the isospin-dependent fusion cross section are needed in order to classify the roles of isospin in fusion dynamics.

\subsection{Interpretability of the model}

\begin{figure*}[hbt!]
    \centering
    \includegraphics[width=\linewidth]{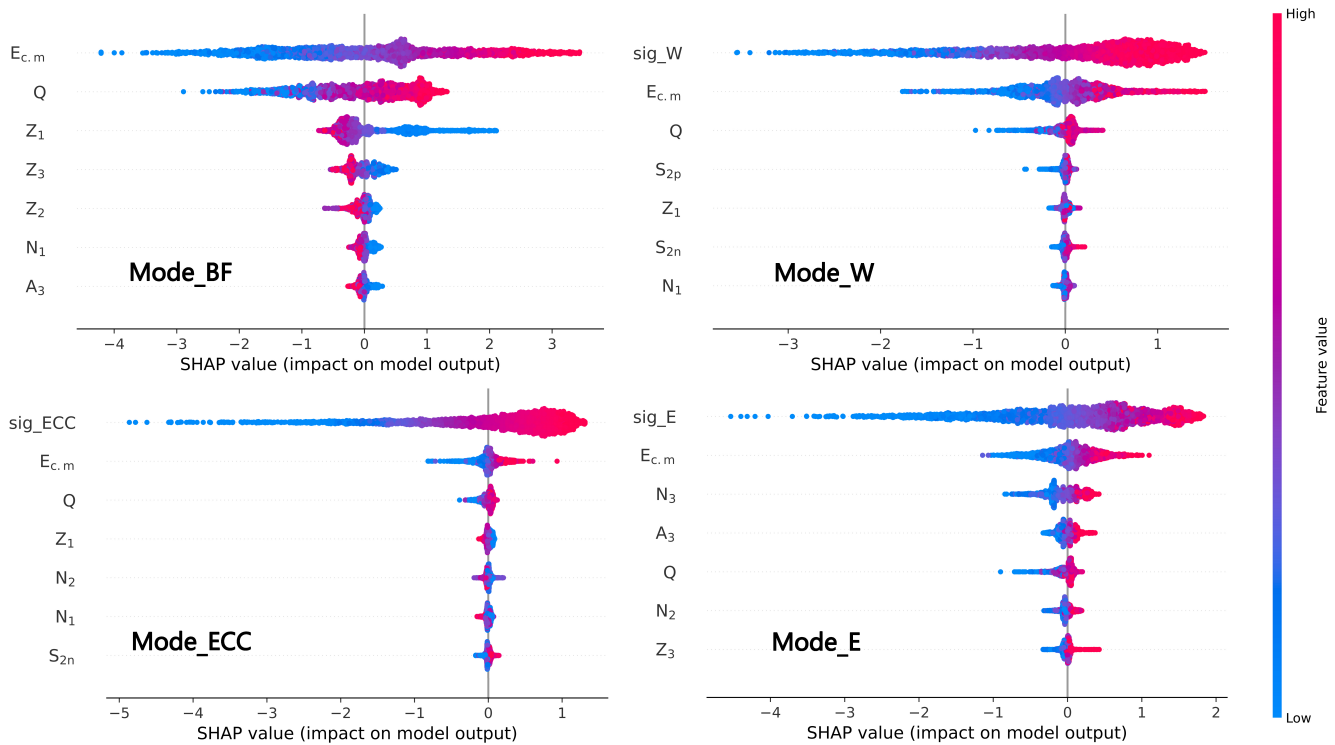}
    \caption{(Color online) Importance ranking for the input features obtained with the SHAP package. Each row represents a feature, and the x-axis is the SHAP value, which shows the importance of a feature for a particular prediction. Each point represents a reaction, and the color represents the feature value (with red being high and blue being low). }
    \label{fig:7}
\end{figure*}

As a decision tree based algorithm, LightGBM has excellent interpretability. This is important because one expects the ML algorithm to not only perform well in refining the theoretical fusion cross section model, but also to provide some fundamental physics that the theoretical fusion cross section model does not have. Understanding what happens when the ML algorithm makes predictions can help us further improve our knowledge of the relationship between the input feature quantities and the predicted values. One possible way to understand how LightGBM provides specific prediction is to find the most important features that drive the model. To do this, one of the most popular feature attribution methods, SHapley Additive Prediction (SHAP)\cite{lundberg2017unified}, is applied to obtain the importance ranking of input features, as displayed in Fig \ref{fig:7}. The top is the most important feature, while the bottom is the least relevant feature for the prediction of the fusion cross section in each mode. It is seen that physical-informed features (sig\_ECC, sig\_W, and sig\_E) in Mode\_W, Mode\_ECC, and Mode\_E are ranked in the top, and their SHAP values are significantly larger than others. Besides these physical-informed features, the collision center-of-mass energy $E_{\rm c.m}$ and fusion Q-value are also ranked in the top-five. It is well known that these two quantities are essential to the process of heavy-ion fusion reaction. In addition, it can be seen that the neutron number (N$_3$) and mass number (A$_3$) of the compound nucleus also exhibit high importance, this indicates these two quantities are strongly related to the fusion cross section, which can be further considered in modeling the fusion cross section.

\section{Summary}

To summarize, the underlying basic quantities and the physical-informed quantities are fed to LightGBM to predict the cross section for heavy-ion fusion reaction. The physical-informed quantities used in this work include the fusion cross sections calculated with the empirical coupled channel (ECC) model and Wong formula, as well as a simplified quantity Z$_1$Z$_2$/E$_{c.m}$. It is found that, by using only basic quantities, LightGBM can reproduce experimental data of cross section within a factor of 10$^{0.129}$=1.35, which is better than 10$^{0.154}$=1.43 obtained from the coupled channel model. When the physical-informed quantities are included in the input feature, the performance of LightGBM can be significantly improved. The MAE on the test set which consists of 118 data points from 8 reaction systems is about 0.66 by only using basic quantities as the input, whereas it is reduced to about 0.2 if the physical-informed quantities are included. In addition, the trend of the excitation function can be reproduced by LightGBM when the input feature includes the physical-informed quantities. All together, our study demonstrates the importance of physical information in predicting fusion cross section
of heavy-ion reaction with machine learning algorithm.

\section{Acknowledgement}

The authors are grateful to the C3S2 computing center in Huzhou University for calculation support. The work is supported in part by the National Natural Science Foundation of China (Nos. U2032145, 12075327, 12335008), Fundamental Research
Funds for the Central Universities, Sun Yat-sen University under Grant No. 23lgbj003, and Guangdong Major
Project of Basic and Applied Basic Research under Grant
No. 2021B0301030006.

\bibliography{arXiv}
\bibliographystyle{elsarticle-num}

\end{document}